\documentclass[hyper]{JHEP} 

\usepackage{epsfig}

\newcommand\fverb{\setbox\pippobox=\hbox\bgroup\verb}

\newcommand\fverbdo{\egroup\medskip\noindent%

            \fbox{\unhbox\pippobox}\ }

\newcommand\fverbit{\egroup\item[\fbox{\unhbox\pippobox}]}

\newbox\pippobox

\title{Note About Weyl Invariant  Ho\v{r}ava-Lifshitz Gravity}
\author{J. Kluso\v{n}\\
Department of
Theoretical Physics and Astrophysics\\
Faculty of Science, Masaryk University\\
Kotl\'{a}\v{r}sk\'{a} 2, 611 37, Brno\\
Czech Republic\\
E-mail: \email{klu@physics.muni.cz}}
\preprint{}

 \abstract{We construct Ho\v{r}ava-Lifshitz gravities that are invariant
 under anisotropic Weyl scaling. This construction is based
 on an extension of the group of symmetries of healthy
 extended Ho\v{r}ava-Lifshitz gravity
 and
 RFDiff invariant Ho\v{r}ava-Lifshitz
 gravity. We find their Hamiltonian
 formulation and determine their
 constraint structure.}
\def\bnabla{\bar{\nabla}}
\def\bN{\bar{N}}
\def\bGamma{\bar{\Gamma}}
\def\baK{\bar{K}}

\def\bg{\bar{g}}

\def\be{\begin{equation}}
\def\bD{\mathbf{D}}
\def\ee{\end{equation}}

\def\bea{\begin{eqnarray}}

\def\eea{\end{eqnarray}}

\def\bR{\bar{R}}

\def\hK{\hat{K}}

\def\mH{\mathcal{H}}

\def\bz{\mathbf{z}}

\def\partt{\partial_t}

\def\bx{\mathbf{x}}
\def\by{\mathbf{y}}
\def \mD{\mathcal{D}}

\newcommand{\mA}{\mathcal{A}}

\newcommand{\mG}{\mathcal{G}}

\def\mV{\mathcal{V}}

\newcommand{\bT}{\mathbf{T}}

\def\pb #1{\left\{#1\right\}}

\begin{document}
\section{Introduction and Summary}\label{first}
In 2009 Petr Ho\v{r}ava formulated new
proposal of quantum theory of gravity
(now known as Ho\v{r}ava-Lifshitz
gravity (HL gravity) that is power
counting renormalizable
\cite{Horava:2009uw,Horava:2008ih,Horava:2008jf}
that is also expected that it reduces
do General Relativity in the infrared
(IR) limit \footnote{For review and
extensive list of references, see
\cite{Horava:2011gd,Padilla:2010ge,Mukohyama:2010xz,Weinfurtner:2010hz}.}.
The HL gravity is based on
 an idea that
the Lorentz symmetry is restored in IR
limit of given theory while it is
absent in its  high energy regime. For
that reason
  Ho\v{r}ava considered
systems whose scaling at short
distances exhibits a strong anisotropy
between space and time,
\begin{equation}
\bx' =l \bx \ , \quad t' =l^{z} t \ .
\end{equation}
In $(D+1)$ dimensional space-time in
order to have power counting
renormalizable theory  requires that
$z\geq D$. It turns out however that
the symmetry group of given theory is
reduced from the full diffeomorphism
invariance of General Relativity  to
the foliation preserving diffeomorphism
\begin{equation}\label{fpdi}
x'^i=x^i+\zeta^ i(t,\bx) \ , \quad
t'=t+f(t) \ .
\end{equation}
Due to the fact that the diffeomorphism
is restricted (\ref{fpdi})  one more
degree of freedom appears that is a
spin$-0$ graviton. It turns out that
the existence of this mode could be
dangerous since it has to decouple
 in the IR regime, in order to
be consistent with observations, for
detailed discussion, see
\cite{Visser:2011mf} and
\cite{Bogdanos:2009uj}. More precisely,
for further discussion it is necessary
to stress that there exists two main
versions of the HL gravity:
projectable, where the lapse function
$N$ depends on $N(t)$ only. This
presumption has a fundamental
consequence for the formulation of the
theory since there is no local form of
the Hamiltonian constraint but only the
global one. The fact that this is the
theory where the local Hamiltonian
constraint is absent implies an
existence of an additional scalar mode.
 The second version of
HL gravity is the version where the
projectability condition is not imposed
so that $N=N(\bx,t)$ \footnote{For
another proposal of renormalizable
theory of gravity, see
\cite{Nojiri:2009th,Nojiri:2010tv}.}.
This form of HL gravity was extensively
studied in
\cite{Blas:2009yd,Blas:2009qj,Blas:2009ck,Blas:2010hb,Li:2009bg,Blas:2010hb,
Kluson:2010xx,Kluson:2010nf,Bellorin:2010te,Bellorin:2010je,Kobakhidze:2009zr,Pons:2010ke,Wang:2010uga}.
It was shown  in \cite{Blas:2010hb}
that this version of HL gravity could
really be an interesting candidate for
the quantum theory of gravity without
ghosts and without strong coupling
problem despite its unusual Hamiltonian
structure
\cite{Kluson:2010xx,Kluson:2010nf}.

Recently Ho\v{r}ava and Malby-Thompson
in \cite{Horava:2010zj}
 proposed very interesting way how
to eliminate the spin-0 graviton in the
context of the projectable version of
HL gravity. Their construction is based
on an
 extension of
the foliation preserving diffeomorphism
in such a way that the theory is
invariant under additional
 local $U(1)$ symmetry. The
resulting theory is known  as
non-relativistic  covariant HL gravity
\footnote{This theory was also studied
in
\cite{Greenwald:2010fp,Alexandre:2010cb,Wang:2010wi,Huang:2010ay,Kluson:2010za,Kluson:2010zn}.}.
It was shown in
\cite{Horava:2010zj,daSilva:2010bm}
that the presence of this new symmetry
implies that the spin-0 graviton
becomes non-propagating and the
spectrum of the linear fluctuations
around the background solution
coincides with the fluctuation spectrum
of General Relativity.

We would like to stress that it is
possible to formulate the modification of HL
gravity that contains the correct
number of physical degrees of freedom
without introducing  additional
fields
\cite{Kluson:2011xx}. This model  is
based on the formulation of the HL
gravity with reduced symmetry group
known as
\emph{restricted-foliation-preserving
Diff} (RFDiff)
 HL gravity
\cite{Blas:2010hb,Kluson:2010na}. This
is the theory that is invariant under
following symmetries
\begin{equation}\label{rfdtr}
t'=t+\delta t \ , \quad \delta
t=\mathrm{const} \ , \quad x'^i=
x^i+\zeta^i(\bx,t) \ .
\end{equation}
The characteristic property of given
theory is the absence of the
Hamiltonian constraint
\cite{Kluson:2010na} either global or
local. The  construction presented
in \cite{Kluson:2011xx} was based on
an extension  of RFDiff HL action
by  an
additional term that is function of
scalar curvature and  it is multiplied
by Lagrange multiplier. It turned out that
 the number of
physical degrees of freedom coincides
with the physical  number of degrees
freedom of General Relativity while
the theory possesses all pleasant properties
of HL gravity.

In this paper we present yet another
version of HL gravity that is now
invariant under  anisotropic Weyl
scaling
\cite{Horava:2009uw,Horava:2008ih} for
general form of the potential term and
for general $\lambda$ \footnote{For
another analysis of HL gravities with
anisotropic Weyl invariance, see
\cite{Moon:2010wq}.}.
 Since lapse transforms non-trivially
 under Weyl scaling it seems to be natural
 to consider HL gravity without the projectability
 condition imposed.
For that reason it makes sense to
discuss invariance of the action under
Weyl scaling in case of the healthy
extended HL gravity or in the theory
where the lapse function is absent as
for example RFDiff HL gravity.

It is well known that the HL gravity
that obeys the detailed balance
condition is invariant under Weyl
rescaling for special case
$\lambda=1/3$ for HL gravity in $D=3$
dimensions. However it was quickly
realized that HL gravity with condition
of detailed balance should be
generalized to the more general form of
the potential given as the linear
 combination of spatial metric,
Ricci tensors and its covariant
derivatives. As a result the
anisitropic Weyl invariance is lost
even in the case of $\lambda=1/3$. Then
it turns out that the only way how to
define Weyl invariant HL gravity for
general $\lambda$  is to built it from
 the connection that
is manifestly invariant under the Weyl
transformation. To do this we introduce
additional scalar field that transforms
under Weyl scaling  in such a way that
it compensates the transformation of
the connection. Clearly this new scalar
field has the form of the
St\"{u}ckelberg field
\cite{Ruegg:2003ps}. In fact, Weyl
scaling symmetry can be fixed by
setting this scalar field to be
constant leading to the recovery of the
original theory.

As we argued above we can introduce HL
gravity invariant under Weyl scaling
for two versions of HL gravity: The
healthy extended HL gravity with
spatial dependent lapse function and in
case of RFDiff invariant HL gravity
with lapse function absent. In the
first case the requirement of the
invariance of the action under Weyl
scaling  only introduces an additional
degree of freedom to the theory. We
expect that this new degree of freedom
does not modify physical content of the
theory due to the fact that it is
accompanied by additional gauge
symmetry. On the other hand the more
interesting case occurs in case of
 RFDiff invariant HL gravity. We show
that the resulting theory can be
modified in the similar way as in case
of the non-relativistic covariant HL
gravity \cite{Horava:2010zj}.
Explicitly, we introduce an additional
term to the action that breaks the
manifest Weyl symmetry of the theory.
After this modification the action
becomes invariant under Weyl scaling on
condition that  the parameter of
transformation is covariantly constant.
Then  in order to restore the
invariance for any value of this
parameter  we introduce the gauge field
that transforms in an appropriate way
under Weyl scaling, following similar
procedure performed in
\cite{Horava:2010zj}. Further, the
Hamiltonian analysis of given theory
shows that it has the same  number of
physical degrees of freedom as in case
of General Relativity. On the other
hand we should stress that the
elimination of the scalar mode that is
present in the original version of HL
gravity is due to the fact that the
newly  introduced gauge field acts as
Lagrange multiplier whose existence
leads to the additional constraint in
the theory. However we mean that it is
sometimes  useful to extend theory by
additional symmetry. In particular, we
can hope that the presence of an
additional symmetry could help to find
 the relation between IR limit of
HL gravity and General Relativity even
if this important problem will not be
studied in this paper. In particular
when we formulate the RFDiff HL gravity
invariant under Weyl scaling we have to
introduce additional scalar field $N$
that can be interpreted as the lapse.
Note also that the lapse $N$ and
spatial metric components $g_{ij}$
transform non-trivially under Weyl
scaling which is different from the
case of the $\alpha-$ symmetry in
covariant non-relativistic HL gravity.
We also  mean that Weyl invariant
RFDiff HL gravity could be also
considered as a new example of the
power counting renormalizable theory of
gravity with correct number of degrees
of freedom.

On the other hand problems and open
questions that are well known from the
formulation of the non-relativistic
covariant HL gravity hold in our case
as well. For example, it is not
completely clear how the IR limit of
this  theory  is related to the General
Relativity. In order to properly
address this issue we should carefully
analyze how the Weyl invariant RFDiff
HL gravity couples to matter. In
particular, it would be interesting to
formulate the action for the probe in
the context of  Weyl invariant RFDiff
HL gravity and study its dynamics. We
should also stress one important point
that is common for   non-relativistic
covariant HL gravity, Weyl invariant
RFDiff HL gravity and  HL gravity with
Lagrange multiplier which is the
presence of the second class
constraints with complicated structure
so that we are not able to solve them
in the full generality. Generally it is
well known that is very difficult task
to analyze theories with the second
class constraints and then perform its
quantum mechanical generalization. One
way how to proceed is to implement the
abelian conversion of the second class
constraints
 \cite{Batalin:1991jm} so that the
 resulting theory can be  formulated
 as the
theory with the  first class
constraints. Then we can certainly
apply the powerful BRST quantization of
such a theory at leas in principle. On
the other hand the fact that the
Poisson brackets between the second
class constrains depend on the phase
space variables implies that the
resulting Hamiltonian will contain
infinite number of terms so that it
seems that given procedure is
meaningless in the case of HL gravity.

Let us outline the content of given
paper. In next section (\ref{second})
we formulate healthy extended HL
gravity that is invariant under
anisotropic Weyl scaling. In section
(\ref{third}) we perform its
Hamiltonian analysis. In section
(\ref{fourth}) we introduce RFDiff HL
gravity invariant under Weyl scaling
and we perform its Hamiltonian analysis
in section (\ref{fifth}).

\section{Weyl Invariant Healthy
Extended HL Gravity} \label{second} In
this section we extend the healthy
extended HL gravity so that the
resulting theory is invariant under
anisotropic Weyl scaling for general
value of parameter $\lambda$ and for
general form of the potential term. As
usual we begin our presentation with
 the introduction of the basis
notation.
Let us consider $D+1$ dimensional
manifold $\mathcal{M}$ with the
coordinates $x^\mu \ , \mu=0,\dots,D$
and where $x^\mu=(t,\bx) \ ,
\bx=(x^1,\dots,x^D)$. We presume that
this space-time is endowed with the
metric $\hat{g}_{\mu\nu}(x^\rho)$ with
signature $(-,+,\dots,+)$. Suppose that
$ \mathcal{M}$ can be foliated by a
family of space-like surfaces
$\Sigma_t$ defined by $t=x^0$. Let
$g_{ij}, i,j=1,\dots,D$ denotes the
metric on $\Sigma_t$ with inverse
$g^{ij}$ so that $g_{ij}g^{jk}=
\delta_i^k$. We further introduce the operator
$\nabla_i$ that is covariant derivative
defined with the metric $g_{ij}$.
 We  introduce  the
future-pointing unit normal vector
$n^\mu$ to the surface $\Sigma_t$. In
ADM variables we have
$n^0=\sqrt{-\hat{g}^{00}},
n^i=-\hat{g}^{0i}/\sqrt{-\hat{g}^{
00}}$. We also define  the lapse
function $N=1/\sqrt{-\hat{g}^{00}}$ and
the shift function
$N^i=-\hat{g}^{0i}/\hat{g}^{00}$. In
terms of these variables we write the
components of the metric
$\hat{g}_{\mu\nu}$ as
\begin{eqnarray}
\hat{g}_{00}=-N^2+N_i g^{ij}N_j \ ,
\quad \hat{g}_{0i}=N_i \ , \quad
\hat{g}_{ij}=g_{ij} \ ,
\nonumber \\
\hat{g}^{00}=-\frac{1}{N^2} \ , \quad
\hat{g}^{0i}=\frac{N^i}{N^2} \ , \quad
\hat{g}^{ij}=g^{ij}-\frac{N^i N^j}{N^2}
\ .
\nonumber \\
\end{eqnarray}
We further define the extrinsic
curvature
\begin{equation}
K_{ij}=\frac{1}{2N} (\partial_t
g_{ij}-\nabla_i N_j- \nabla_j N_i) \
\end{equation}
and the generalized de Witt metric
$\mG^{ijkl}$ in the
form
\begin{equation}
\mG^{ijkl}=\frac{1}{2}(g^{ik}g^{jl}+
g^{il}g^{jk})-\lambda g^{ij}g^{kl} \ ,
\end{equation}
where $\lambda$ is real constant. In
this notation the HL action has the
form
\begin{equation}\label{actionNOJI}
S=\frac{1}{\kappa^2}
\int dt d^D\bx
\sqrt{g}N [K_{ij}\mG^{ijkl}K_{kl}-
\mV(g)] \ ,
\end{equation}
where  $\mV(g)$ is a general function
of $g_{ij}$ and its covariant
derivative. The action
(\ref{actionNOJI})
 is invariant under foliation
preserving diffeomorphism
\begin{equation}\label{fpd}
t'-t=f(t) \ , \quad
x'^i-x^i=\xi^i(t,\bx) \ .
\end{equation}
Our goal is to formulate HL gravity
that is invariant under anisotropic
Weyl transformation
defined as
\begin{equation}\label{Weylgen}
N'=\Omega^z N \ , \quad N'_i=\Omega^2
N_i \ , \quad g_{ij}=\Omega^2 g_{ij} \
.
\end{equation}
 We construct this modification of
  HL gravity in the  following way. It is easy to show that
  the component of
the spatial connection
\begin{equation}
\Gamma_{ij}^k=
\frac{1}{2}g^{kl} (\partial_i
g_{lj}+\partial_j
g_{li} -\partial_l
g_{ij}) \
\end{equation}
transforms  under (\ref{Weylgen})  as
\begin{eqnarray}\label{Gammacon}
\Gamma'^k_{ij}=
\Gamma^k_{ij}+ \frac{1}{\Omega}
(\delta^k_i\partial_j
\Omega+\delta^k_j\partial_i
\Omega-g^{kl}
\partial_l \Omega g_{ij}) \ .
\nonumber \\
\end{eqnarray}
Then we find that the extrinsic
curvature $K_{ij}$ transforms under
(\ref{Weylgen}) as
\begin{eqnarray}\label{Kex1}
K'_{ij}=
\Omega^{2-z} K_{ij}+\Omega^{1-z}\nabla_n \Omega g_{ij} \ ,
\nonumber \\
\end{eqnarray}
where we defined
\begin{equation}
\nabla_n X=\frac{1}{N}
(\partial_t X-N^i\partial_i X) \ .
\end{equation}
It is important to stress that the
potential term in HL gravity contains
terms that are constructed from
covariant derivatives and Ricci tensors
of higher order. Then  in order to
formulate HL gravity
 that is invariant under
 (\ref{Weylgen}) for general form of the
potential it is natural to introduce
the connection is manifestly invariant
under (\ref{Weylgen}). For that reason
we introduce scalar field
$\varphi(t,\bx)$ that transforms under
anisotropic Weyl scaling as
\begin{equation}\label{trvarphi}
\varphi'(t,\bx)=\Omega(t,\bx)
\varphi(t,\bx) \
\end{equation}
and define the  connection with bar
\begin{equation}
\bar{\Gamma}_{ij}^k= \Gamma_{ij}^k
-\frac{1}{\varphi}(\delta^k_i\partial_j
\varphi +\delta^k_j\partial_i
\varphi-g^{kl}
\partial_l \varphi g_{ij}) \ .
\end{equation}
Then with the help of (\ref{Gammacon})
 and (\ref{trvarphi})we easily find that
$\bar{\Gamma}_{ij}^k$ is invariant
under (\ref{Weylgen}).
As a result $D-$dimensional Riemann
tensor constructed from connection with
bar is invariant under (\ref{Weylgen})
which also implies that corresponding
Ricci tensor and scalar curvature
transform as
\begin{equation}
\bar{R}'_{ij}=\bar{R}_{ij} \ , \quad
\bar{R}'=\Omega^{-2}\bar{R} \ .
\end{equation}
Let us now discuss following  specific
form of the potential $\mV$
\cite{Sotiriou:2009bx}
\begin{eqnarray}
\mV(g)&=&\zeta^2 g_0+ g_1 R+
\frac{1}{\zeta^2}(g_2 R^2+ g_3
R_{ij}R^{ij})+\nonumber \\
&+&\frac{1}{\zeta^4} (g_4 R^3+g_5 R
R_{ij}R^{ij}+g_6 R^i_j R^j_k R^k_i)+
\nonumber \\
&+&\frac{1}{\zeta^4} [g_7 R \nabla^2 R
+g_8(\nabla_i R_{jk}) (\nabla^i
R^{jk})] \ , \nonumber \\
\end{eqnarray}
where the coupling constants $g_s,
(s=0,1,2,\dots,8)$ are dimensionless.
The relativistic limit in the IR
requires $g_1=-1$ and $\zeta^2= (16\pi
G)^{-2}$. Then in order to find the
potential term that is invariant under
(\ref{Weylgen}) we replace $R_{ij}$
with $\bar{R}_{ij}$ and multiply all
expressions with appropriate powers of
$\varphi$ so that
\begin{eqnarray}
\bar{\mV}(\varphi,g)&=& \zeta^2 g_0+
g_1 \varphi^2\bar{R}+
\frac{1}{\zeta^2}\varphi^4(g_2 \bR^2+
g_3
\bR_{ij}\bR^{ij})+\nonumber \\
&+&\frac{1}{\zeta^4} \varphi^{6}(g_4
\bR^3+g_5 \bR \bR_{ij}\bR^{ij}+g_6
\bR^i_j \bR^j_k \bR^k_i)+
\nonumber \\
&+&\frac{1}{\zeta^4} [g_7 \varphi^2 \bR
\bar{\nabla}_i(\varphi^{2}
g^{ij}\bar{\nabla}_j (\varphi^{2}\bR))
+g_8(\bar{\nabla}_i (\bar
R_{jk})\varphi^{6}
g^{il}g^{jm}g^{kn}\bar{\nabla}_l
\bR_{mn})] \ .   \nonumber \\
\end{eqnarray}

 Let us now construct the kinetic
term that is invariant under Weyl
scaling. To do this we introduce
 extrinsic curvature with bar defined as
\begin{equation}
\bar{K}_{ij}=K_{ij}-\frac{1}{\varphi}
\nabla_n\varphi g_{ij} \
\end{equation}
 that transforms under (\ref{Weylgen})
 as
\begin{eqnarray}
\bar{K}'_{ij}=
 \Omega^{2-z}\bar{K}_{ij}
\nonumber \\
\end{eqnarray}
using (\ref{Kex1}) and the fact that
\begin{equation}
\nabla'_n\varphi'=\Omega^{1-z}\nabla_n\varphi+\varphi
\Omega^{-z} \nabla_n\Omega \ .
\end{equation}
Then with the help of the fact that
under Weyl scaling the generalized de
Witt metric transforms as
\begin{equation}
\mG'_{ijkl}=\Omega^4 \mG_{ijkl} \ ,
\quad \mG'^{ijkl}=\frac{1}{\Omega^4}
\mG^{ijkl} \
\end{equation}
we find
\begin{equation} \bar{K}'_{ij}
\mG'^{ijkl}\bar{K}'_{kl}= \Omega^{-2z}
\bar{K}_{ij} \mG^{ijkl}\bar{K}_{kl} \ .
\end{equation}
Collecting these results we propose
following
 HL action that is invariant under
 anisotropic Weyl transformation
\begin{equation}\label{HLconactiona}
S=\frac{1}{\kappa^2} \int dt d^D\bx
\sqrt{g}N \varphi^{-(z+D)}
[\varphi^{2z}\bar{K}_{ij}\mG^{ijkl}\bar{K}_{kl}
-\bar{\mV}(g,\phi)] \ .
\end{equation}
Let us now discuss the spatial
dependence of the  lapse function $N$.
Since the theory is invariant under
anisitropic Weyl transformation
 with parameter
that depends on space and time
coordinates it is natural to presume
that $N$ depends on $\bx$ and $t$ as
well. Then the only way  how to find
consistent HL theory where the lapse is
spatial dependent is to consider its
healthy extended  form that contains
$D-$dimensional vector $a_i$ defined as
\begin{equation}
a_i=\frac{\partial_i N}{N} \ .
\end{equation}
The   healthy extended HL gravity
contains terms that depend on $a_i$ and
that are invariant under spatial
diffeomorphism. We denote this
contribution by introducing an
additional potential $\mV_a(a,g,R)$.
Note also that  $a_i$ transforms under
(\ref{Weylgen})
 as
\begin{equation}
a'_i(\bx,t)=a_i(\bx,t)+\frac{z}{\Omega}\partial_i\Omega(\bx,t) \ .
\end{equation}
We see that in order to find the
healthy extended HL gravity that is
invariant under (\ref{Weylgen}) we have
to  replace  $a_i$ with $a_i\rightarrow
\bar{a}_i\equiv
a_i-\frac{z}{\varphi}\partial_i\varphi
$ and include appropriate factors of
$\varphi$ in corresponding expressions.
In summary we propose following healthy
extended HL gravity action that is
invariant under anisotropic Weyl
transformation
\begin{equation}\label{HLconaction}
S=\frac{1}{\kappa^2} \int dt d^D\bx
\sqrt{g}N \varphi^{-(z+D)}
[\varphi^{2z}\bar{K}_{ij}\mG^{ijkl}\bar{K}_{kl}
-\bar{\mV}(g,\varphi)-\bar{\mV}_a(\bar{a},g,
\bR,\varphi)] \ .
\end{equation}
\section{Hamiltonian Analysis of
Weyl Invariant Healthy Extended HL
Gravity}\label{third} In this section
we
 perform  the Hamiltonian
analysis of the action
(\ref{HLconaction}). As a result of
this analysis we will be able to
determine the number of physical
degrees of freedom.

As the first step we determine the momenta conjugate to
 $N,N^i,g_{ij}$ and $\varphi$ from (\ref{HLconaction})
 %
\begin{eqnarray}\label{defmom}
p_N(\bx)&=&\frac{\delta
S}{\delta
\partt N(\bx)}\approx 0 \ , \quad
p^i(\bx)=\frac{\delta
S}{\delta
\partt N_i(\bx)} \approx 0 \ , \quad
\nonumber \\
\pi^{ij}(\bx)&=&\frac{\delta S}{\delta
\partt
g_{ij}(\bx)}=\frac{1}{\kappa^2}
\sqrt{g} \varphi^{-(z+D)}\varphi^{2z}\mG^{ijkl}\bar{K}_{kl}
\ , \nonumber \\
 p_{\varphi}(\bx)&=&\frac{\delta
S}{\delta
\partial_t \varphi(\bx)}
=-\frac{2}{\kappa^2}
\sqrt{g}\varphi^{-(z+D)}\varphi^{2z-1}
g_{ij}\mG^{ijkl}\bar{K}_{kl} \ .
\nonumber \\
\end{eqnarray}
As usual $p_N,p^i$ are primary
constraints of the theory. Further,
using relations given above we find an
additional primary constraint
\begin{eqnarray}
\mD\equiv 2g_{ij}\pi^{ji}+
p_\varphi\varphi\approx 0
\nonumber \\
\end{eqnarray}
and the bare Hamiltonian
\begin{eqnarray}
H&=&\int d^D\bx (N\mH_T+N_i \mH^i) \ ,  \nonumber \\
\mH^i&=&p_\varphi
g^{ij}\nabla_j\varphi-2 \nabla_j
\pi^{ji} \ ,
\nonumber \\
\mH_T
&=&\frac{\kappa^2}{\sqrt{g}}\varphi^{(z+D)}\frac{1}{\varphi^{2z}}
\pi^{ij}\mG_{ijkl}\pi^{kl}+\frac{1}{\kappa^2}\sqrt{g}\varphi^{-(z+D)}(
\bar{\mV}(g,\varphi)+\bar{\mV}_a(\bar{a},g,\bar{R},\varphi))\equiv \nonumber \\
&\equiv
&\bar{\mH}_T+\frac{1}{\kappa^2}\sqrt{g}\varphi^{-(z+D)}
\bar{\mV}_a(\bar{a},g,\bR,\varphi))\ ,\nonumber \\
\end{eqnarray}
where $\bar{\mH}_T$ coincides with the Hamiltonian
constraint of the HL gravity without its healthy
extension.
 Following
 the
 standard formalism of
constraint system we introduce the
extended Hamiltonian in the form
\begin{eqnarray}
H_T=\int d^D\bx (N(\bar{\mH_T}+\frac{1}{\kappa^2}
\sqrt{g}\varphi^{-(z+D)}\bar{\mV}_a)
+N_i\mH^i+
\lambda_N p_N+\lambda_i p^i+
\lambda^{\mD}\mD)\ .  \nonumber \\
\end{eqnarray}
Now we proceed to the analysis of the
stability of the primary constraints.
As usual the preservation of the primary
constraints $p_i(\bx)\approx 0$
imply the secondary constraints
\begin{equation}
\mH^i(\bx)\approx 0 \ .
\end{equation}
It is convenient to introduce the
following slightly modified
smeared form of this constraint
\begin{equation}
\bT_S(\xi)=\int d^D\bx (\xi^i(\bx)\mH_i(\bx)+
\xi^i(\bx)\partial_i N(\bx)p_N(\bx)) \ .
\end{equation}
Note that the additional term in $\bT_S$
 is proportional to the primary constraint
$p_N(\bx)\approx 0$.
Now we analyze the requirement of
the preservation of the primary
 constraint $ \Theta_1(\bx)\equiv p_N(\bx)\approx 0$
 during the time evolution of the system. Explicitly,
 the time evolution of this constraint
 is governed by following equation
 \begin{eqnarray}\label{parTheta1}
\partial_t \Theta_1(\bx)&=&
\pb{\Theta_1(\bx),H_T}=
-\tilde{\mH_T}(\bx)-\frac{1}{\kappa^2}
\sqrt{g}\varphi^{-(z+D)}
\bar{\mV}_a+\nonumber \\
&+&
\frac{1}{N}\partial_i\left(N\sqrt{g}\varphi^{-(z+D)}\frac{\delta
\bar{\mV}_a}{\delta
\bar{a}_i}\right)(\bx)\equiv
-\Theta_2(\bx)\approx 0 \ ,
\nonumber \\
\end{eqnarray}
where $\Theta_2$ is the secondary
constraint.  Now we analyze the
condition of the stability of the
primary constraint $\mD$. It turns out
to be convenient to introduce the
smeared form of the constraint $\mD$
defined as
\begin{equation}
\bD(\omega)=\int d^D\bx \omega
(\bx)(\mD(\bx)+zN(\bx)p_N(\bx)) \ ,
\end{equation}
that has following Poisson bracket
\begin{eqnarray}
\pb{\bD(\omega),g_{ij}}&=& -2\omega
g_{ij} \ , \quad
\pb{\bD(\omega),g^{ij}}= 2\omega g^{ij}
\ , \nonumber \\
\pb{\bD(\omega),p^{ij}}&=& 2\omega
\pi^{ij} \ , \quad
\pb{\bD(\omega),\varphi}=
-\omega \varphi \ , \nonumber \\
\pb{\bD(\omega),p_\varphi}&=&\omega
p_\varphi  \ , \quad
 \pb{\bD(\omega),N}=-z\omega N \ , \nonumber \\
\pb{\bD(\omega),
a_i}&=&-z\partial_i\omega \ , \quad
\pb{\bD(\omega),\bar{a}_i}=0 \ . \nonumber \\
\end{eqnarray}
Then we find
\begin{eqnarray}
\pb{\bD(\omega),\bar{\Gamma}_{ij}^k}
=0\nonumber \\
\end{eqnarray}
and consequently
\begin{eqnarray}
\pb{\bD(\omega),\bar{R}^k_{ijm}}=0 \ ,
\quad \pb{\bD(\omega),\bar{R}_{ij}}= 0
\ , \quad
\pb{\bD(\omega),\bar{R}}=-2\omega
\bar{R} \ .
\nonumber \\
\end{eqnarray}
Using these results it is easy to see
that
\begin{equation}\label{pbbDHt}
\pb{\bD(\omega), \bar{\mV}}=0 \ , \quad
\pb{\bD(\omega),\bar{\mV}_a}=0 \ ,
\quad \pb{\bD(\omega),\mH_T}=z\omega
\mH_T \ .
\end{equation}
In order to analyze the Poisson bracket
including $\bT_S(\xi)$ we  need
following Poisson brackets
 \begin{eqnarray}\label{pbBTS}
 \pb{\bT_S(\xi),g_{ij}}&=&
 -\xi^k\partial_k
 g_{ij}-
 \partial_i \xi^k
  g_{kj} -g_{ik}\partial_j\xi^k \ ,
 \nonumber \\
\pb{\bT_S(\xi),p^{ij}}&=& -\partial_k
p^{ij} \xi^k-p^{ij}\partial_k
\xi^k+\partial_k \xi^ip^{kj} +p^{ik}
\partial_k \xi^j \ , \nonumber \\
\pb{\bT_S(\xi),a_i}
&=&-\xi^j\partial_j a_i-\partial_i
\xi^j
 a_j \ . \nonumber \\
 \end{eqnarray}
Then we easily find
\begin{eqnarray}\label{pbBTS1}
\pb{\bT_S(\xi),\mH_T}&=&
-\xi^k\partial_k \mH_T-
\mH_T\partial_k \xi^k \ ,
\nonumber \\
\pb{\bT_S(\xi),\mV_a(g)}&=& -\partial_i
V\xi^i \  .
\nonumber \\
\pb{\bT_S(\xi),\mD}&=& -\partial_k
\mD\xi^k-\mD \partial_k\xi^k \ .
\nonumber \\
\end{eqnarray}
The last Poisson bracket has following
smeared form
\begin{equation}\label{pbbDbTS}
\pb{\bD(\omega),\bT_S(\xi)}=
\bD(-\partial_k \omega \xi^k) \ .
\end{equation}
This result together with
the last equation in (\ref{pbbDHt})
implies that the constraint $\bD$ is
preserved during the time evolution
of the system.

Including the secondary constraint $\Theta_2$
into the definition of the Hamiltonian we
find that the total Hamiltonian takes the form
\begin{equation}\label{Hamhealtcon}
H=\int d^D\bx(
N(\bar{\mH}_T+\frac{1}{\kappa^2}
\sqrt{g}\varphi^{-(z+D)}
\bar{\mV}_a)+\lambda_i
p^i+v^\alpha\Theta_\alpha)
+\bT_S(N^i)+\bD(\lambda^\mD) \ ,
\end{equation}
where $v^\alpha$ are Lagrange multipliers related
to the  constraints $\Theta_\alpha $.

As the next step we have to check the stability
of the secondary constraints $\Theta_2(\bx)\approx 0 \ ,
\bT_S(\xi)\approx 0$. Using
(\ref{pbBTS}) and (\ref{pbBTS1}) we find
\begin{eqnarray}\label{bTSTheta}
\pb{\bT_S(\xi),\Theta_\alpha(\bx)}=
-\partial_k \Theta_\alpha
(\bx)\xi^k(\bx)-
\Theta_\alpha(\bx)\partial_k \xi^k(\bx) \ .
\nonumber \\
\end{eqnarray}
With the help of this  result and also
(\ref{pbbDbTS}) we find that the
constraint $\bT_S(\xi)\approx 0$ is
preserved during the time evolution of
the system.
Now we proceed to the analysis of the
stability of the constraints
$\Theta_{1,2}$.  First of all it is
easy to see that
\begin{equation}
\pb{\bD(\omega),\Theta_1}=\omega\Theta_1
\ , \quad
\pb{\bD(\omega),\Theta_2}=z\omega
\Theta_2 \ .
\end{equation}
Then with the help of (\ref{bTSTheta})
we find that the Poisson brackets of
$\Theta's$ with $\bD$ and with $\bT_S$
vanish on the constraints surface. To proceed
further  we calculate
following Poisson bracket
\begin{eqnarray}
& &\pb{\Theta_1(\bx),\Theta_2(\by)}\equiv  \triangle_{12}(\bx,\by)
=\nonumber \\
& &=-\frac{1}{N}\partial_{y^i}
\left(\sqrt{g} \frac{\delta^2 V}{\delta \bar{a}_i(\by)
\delta \bar{a}_j(\by)}\left(\bar{a}_j(\by)
\delta(\bx-\by)-\partial_{y^j}
\delta(\bx-\by)\right) \right)
 \ .  \nonumber \\
\end{eqnarray}
Collecting  these results we find that
the time evolution of the constraint
$\Theta_1(\bx)$ is equal to
\begin{eqnarray}
\partial_t \Theta_1(\bx)&=&
\pb{\Theta_1(\bx),H_T}\approx
\int d^D\by v_2\triangle_{12}(\bx,\by) \ .
\nonumber \\
\end{eqnarray}
Clearly  $\partial_t \Theta_1\approx 0$
for  $v_2=0$. In the
same way we determine the time evolution of
the constraint $\Theta_2(\bx)\approx 0$
\begin{eqnarray}\label{Theta2con}
\partial_t \Theta_2(\bx)&=&\pb{\Theta_2(\bx),H_T}\approx
\nonumber \\
&\approx &\int d^D\by \left(
N\pb{\Theta_2(\bx),\bar{\mH}_T(\by)+
\frac{1}{\kappa^2}\varphi^{-(z+D)}
\sqrt{g}\bar{\mV}_a(\by)}
-v_1\triangle_{12}(\by,\bx) \right)=0 \
\nonumber \\
\end{eqnarray}
using $v_2=0$. From (\ref{Theta2con})
we see that the requirement that
$\partial_t \Theta_2(\bx)=0$ fixes the
value of the Lagrange multiplier $v_1$.
Equivalently, $\Theta's$ are the second
class constrains that according to
standard analysis have to vanish
strongly.
As the result of this analysis
we find  following extended Hamiltonian
\begin{equation}
H_T=H+\bT_S(N^i)+\bD(\lambda^\mD)+\int
d^D\by \lambda_i p^i \ ,
\end{equation}
where
\begin{equation}\label{Hhealthy}
H=\int d^D\bx N\left(
\bar{\mH}_T+\frac{1}{\kappa^2}
\sqrt{g}\varphi^{-(z+D)}\bar{\mV}_a\right)
 \ .
\end{equation}
As we argued above the constraints
$\Theta's$ are the second class
constraints that should be solved for
the canonical pair $p_N,N$, at least in
principle.
At the same time it  is necessary to
replace the Poisson brackets between
phase space variables
$(g_{ij},\pi^{ij})$  with the Dirac
brackets
\begin{eqnarray}
\pb{F(g,p),G(g,p)}_D&=&
\pb{F(g,p),G(g,p)}-\nonumber \\
&-&\int d^D\bx d^D\by \pb{F(q,p),
\Theta_\alpha(\bx)}
\triangle^{\alpha\beta}(\bx,\by)\pb{\Theta_\beta(\by),G(p,q)}
\ ,
\nonumber \\
\end{eqnarray}
where $\triangle^{\alpha\beta}(\bx,\by)$ is inverse of $\triangle_{\alpha\beta}(\bx,\by)$
in a sense
\begin{equation}
\int d^D\bz \triangle_{\alpha\beta}(\bx,\bz)
\triangle^{\beta\gamma}(\bz,\by)
=\delta_\alpha^\gamma\delta(\bx-\by) \ .
\end{equation}
However due to the fact that
the Poisson brackets
between $g_{ij},p^{ij}$ and $\Theta_1$ vanish
we find that the Dirac brackets between
canonical variables $g_{ij},p^{ij}$
 coincide with the
Poisson brackets.
Even if we are not able to solve the
constraint $\Theta_2$ explicitly  we
can still   determine the number of
physical degrees of freedom knowing the
constraint structure of the theory
\cite{Henneaux:1992ig}. We have
$N_{p.s.d.f.}=D(D+1)+2D+4$ phase space
variables $g_{ij},
\pi^{ij},N,p_N,N_i,p^
i,\varphi,p_\varphi$. On the other hand
we have $N_{f.c.c.}=2D+1$ first class
constraints $p^i\approx 0 \ ,
\mH_i\approx 0 \ , \mD\approx 0 $ and
$N_{s.c.c.}=2$ the second class
constraints $\Theta_{1,2}\approx 0$.
Then the number of physical degrees of
freedom is \cite{Henneaux:1992ig}
\begin{equation}
N_{p.d.f}= N_{p.s.d.f.}-2N_{f.c.c.}-
N_{s.c.c}=(D^2-D-2)+2 \ ,
\end{equation}
 where the expression in parenthesis
corresponds to the number of physical
degrees of freedom of general
relativity and where the factor $2$
corresponds to the additional scalar
mode whose presence is the general
property of the healthy extended HL
gravity. On the other hand the
non-relativistic covariant HL gravity
proposal completely eliminates this
mode. In the next section we formulate
the Weyl invariant  RFDiff HL gravity
where
this scalar mode is also eliminated.
\section{Weyl Invariant  RFDiff HL Gravity}
\label{fourth}
 The RFDiff invariant HL
gravity was introduced in
\cite{Blas:2010hb} and further studied
in
\cite{Kluson:2010na,Kluson:2010zn,Kluson:2011xx}.
This is the version of the HL gravity
that is invariant under restricted
foliation preserving diffeomorphism
\begin{equation}\label{rfd}
t'=t+\delta t \ , \quad  \delta
t=\mathrm{const} \ , \quad  x'^i=x^i+
\zeta^i(t,\bx) \ .
\end{equation}
The simplest form of RFDiff invariant
Ho\v{r}ava-Lifshitz gravity takes the
form \cite{Kluson:2010na}
\begin{equation}\label{RFDiffaction}
S=\frac{1}{\kappa^2} \int dt d^D\bx
\sqrt{g}(\hK_{ij}
\mG^{ijkl}\hK_{kl}-\mV(g)) \ ,
\end{equation}
where we introduced modified extrinsic
curvature
\begin{equation}
\hK_{ij}=\frac{1}{2}(\partial_t g_{ij}
-\nabla_i N_j-\nabla_j N_i) \
\end{equation}
that differs from the standard
extrinsic curvature by absence of the
lapse $N$. In fact, the absence of the
lapse is the general property of RFDiff invariant
HL gravity even if it is possible to
consider more general case \cite{Blas:2010hb}
where however $N$ transforms as scalar
 under (\ref{rfd}).
Now in order to formulate RFDiff  HL
gravity that is invariant under
anisotropic Weyl transformation we
proceed in the slightly different way
than in section (\ref{second}). Explicitly,
we replace $g_{ij}$ and $N_i$ with
corresponding  variables with bar
defined as
\begin{equation}
\bg_{ij}=\frac{1}{N^{2/z}}g_{ij} \ ,
\quad
\bN_i=\frac{1}{N^{2/z}}N_i \ ,
\end{equation}
where $N$ transforms as  scalar under
(\ref{rfd}).
 Clearly $\bg$ and $\bN_i$ are invariant under
 transformations
\begin{equation}\label{Weyl2}
N'(\bx,t)=\Omega^z(\bx,t)N(\bx,t) \ ,
\quad g'_{ij}(\bx,t)=\Omega^2
(\bx,t)g_{ij}(\bx,t) \ , \quad
N'_i(\bx,t)=\Omega^2 N_i(\bx,t) \ .
\end{equation}
As a result   any  theory constructed
from $\bg$ and $\bN_i$ is invariant
under the transformation (\ref{Weyl2}).
On the other hand it is useful to find
relation between quantities with and
without bar. Explicitly
\begin{eqnarray}
\bGamma_{ij}^k&=& \frac{1}{2}
\bg^{kl}(\partial_i \bg_{lj}+
\partial_j \bg_{li}-\partial_l \bg_{ij})=
\nonumber \\
&=&
\Gamma_{ij}^k-
\frac{1}{z}(a_i\delta_j^k+a_j
\delta_i^k-g^{kl} a_l g_{ij}) \ , \quad
a_i=\frac{\partial_iN}{N} \nonumber
\\
\end{eqnarray}
and
\begin{eqnarray}
\baK_{ij}= \frac{1}{2}(\partial_t
\bg_{ij} -\bar{\nabla}_i \bN_j-
\bnabla_j \bN_i)
=\frac{1}{N^{2/z}}(
\hK_{ij}-\frac{1}{z}\nabla_n N g_{ij})
\ ,
\nonumber \\
\end{eqnarray}
where again $\nabla_n X=\frac{1}{N}
(\partial_t X-N^ i \partial_i X)$.
Then it is easy to see
that  RFDiff HL gravity defined by the
action
\begin{eqnarray}
S&=&\frac{1}{\kappa^2} \int  dt d^D \bx
 \sqrt{\bg}(\baK_{ij}\bar{\mG}^{ijkl}
\baK_{kl}-\mV (\bg, \bR))=
\nonumber \\
&=&\frac{1}{\kappa^2} \int dt d^D \bx
\sqrt{g}N^{-D/z}\left[(\hK_{ij}-\frac{1}{z}\nabla_n
N g_{ij}) \mG^{ijkl}(
\hK_{kl}-\frac{1}{z}\nabla_n N g_{kl})
-\mV (\bg, \bR)\right]
\nonumber \\
\end{eqnarray}
is invariant under (\ref{Weyl2}).
Apparently we could expect this result
since $N$  cannot change the physical
content of the theory as follows from
the way we introduced it into the
action. At this place we would like to
stress the analogy with the similar
situation in non-relativistic covariant
HL gravity. We argued in
\cite{Kluson:2010zn,Kluson:2010za} that
the  field $\nu$ that is
present in the covariant
non-relativistic HL gravity should be
interpreted as St\"{u}ckelberg field.
Note that we could give the same interpretation
to the field $N$ introduced above.

The idea how to make   the mode $N$
non-trivial  is
 to add to the action an additional
term that breaks the manifest Weyl
symmetry of the action. To do this we
proceed in the similar way as in case
of non-relativistic covariant HL
gravity and  consider following term
\begin{equation}
S_{A}=\frac{1}{\kappa^2} \int dt d^D\bx
\sqrt{\bg}\mG(\bR)\nabla_n N \ ,
\end{equation}
where $\mG$ is the scalar function that
depends on $\bR$ only. Note that under
Weyl transformation $S_A$  transforms
as
\begin{equation}
S'_{A}= \frac{1}{\kappa^2} \int d^D\bx
\sqrt{\bg}\mG(\bR)(\nabla_n N
+\frac{z}{\Omega}(\partial_t\Omega-N^i\partial_i
\Omega)) \ .
\end{equation}
In other words the theory is not
invariant under general local Weyl
transformation but only under
transformation with covariantly
constant parameter:
\begin{equation}
\partial_t\Omega-N^i\partial_i \Omega=0 \ .
\end{equation}
We see that this condition coincides
with the restriction of the parameter
of the $\alpha-$transformation in case
of non-relativistic HL gravity
\cite{Horava:2010zj}. Then following
this paper we make the theory invariant
for general $\Omega$ when  we introduce
the gauge field $\mA$ into the action
 $S_{A}$ so that
\begin{equation}\label{addterm}
S_A=\frac{1}{\kappa^2} \int d^D\bx
\sqrt{\bg}\mG(\bR)(\nabla_n N-\mA)
\end{equation}
and  demand that $\mA$ transforms under
Weyl scaling as
\begin{equation}
\mA'(t,\bx)=\mA(t,\bx)+\frac{z}{\Omega(t,\bx)}
(\partial_t\Omega(t,\bx)-N^i\partial_i\Omega(t,\bx))
\ .
\end{equation}
Then we see that (\ref{addterm})  is
invariant under  Weyl transformation
for general values of the parameter
$\Omega$. The presence of the field
$\mA$ has  crucial impact on the
physical content of given theory as
will be seen from its Hamiltonian
analysis.
\section{Hamiltonian Formalism for
Weyl Invariant RFDiff  HL
gravity}\label{fifth}
 For reader's convenience  we
again write  Weyl invariant
 RFDiff HL gravity action
\begin{eqnarray}\label{actHam}
S&=&\frac{1}{\kappa^2} \int d^D \bx dt
\sqrt{g}N^{-D/z}\left((\hK_{ij}-\frac{1}{z}\nabla_n
N g_{ij}) \mG^{ijkl}(
\hK_{kl}-\frac{1}{z}\nabla_n N
g_{kl})\right.
-\nonumber \\
&-& \left. \mV (\bg, \bR)+
\mG(\bR)(\nabla_n
N-\mA)\right) \ . \nonumber \\
\end{eqnarray}
As the first step we determine momenta
from (\ref{actHam})
\begin{eqnarray}
\pi^{ij}&=&\frac{\delta S}{\delta
\partial_t g_{ij}}=\frac{1}{\kappa^2}
\sqrt{g}N^{-D/z}\mG^{ijkl}(\hK_{kl}-
\frac{1}{z}\nabla_n N g_{kl}) \ , \quad  p^i=
\frac{\delta S}{\delta \partial_t
N_i}\approx 0 \ , \nonumber \\
p_N&=&\frac{\delta S}{\delta\partial_t
N}=-\frac{2}{\kappa^2} \sqrt{g}N^{-D/z}
\frac{1}{zN}
g_{ij}\mG^{ijkl}(\hK_{kl}-\frac{1}{z}\nabla_n
N g_{kl})+ \nonumber
\\
&+& \frac{1}{\kappa^2}
\sqrt{g}N^{-D/z-1}\mG(\bR) \ , \quad  p_\mA\approx 0 \ .  \nonumber \\
\end{eqnarray}
We see  that this theory possesses
$D+2$  primary constraints:
\begin{eqnarray}\label{primcon2}
& & p^i \approx 0 \ , \quad
p_\mA\approx 0 \ , \nonumber \\
& &\mD\equiv zp_N
N+2\pi^{ij}g_{ji}-\frac{1}{\kappa^2}
\sqrt{g}N^{-D/z}\mG(\bR)\equiv \mD_0
-\frac{1}{\kappa^2}
\sqrt{g}N^{-D/z}\mG(\bR) \approx 0 \ .
 \nonumber \\
\end{eqnarray}
Following standard procedure we find
Hamiltonian
\begin{eqnarray}
H_T&=&\int d^D\bx (\mH_T+N_i
\mH^i+v_\mA p_\mA+
v_i p^i+v^\mD \mD) \ ,  \nonumber \\
\mH^i&=& p_N g^{ij}\nabla_j N-2
\nabla_k \pi^{ki} \ ,
\nonumber \\
\mH_T &=&
\frac{\kappa^2}{\sqrt{g}}N^{D/z}
\pi^{ij}\mG_{ijkl}\pi^{kl}
+\frac{1}{\kappa^2}\sqrt{g} N^{-D/z}
\mV(g,\phi)+\frac{1}{\kappa^2}
\sqrt{g}N^{-D/z}\mG(\bR)\mA \ , \nonumber \\
\end{eqnarray}
where $v_\mA,v_i,v^\mD$ are Lagrange
multipliers corresponding to the
primary constraints (\ref{primcon2}).
Now we proceed to the analysis of the
preservation of these constraints. In
case of $p_\mA,p^i$ we find
\begin{eqnarray}
\partial_t p_\mA&=&\pb{p_\mA,H_T}=
-\frac{1}{\kappa^2}
\sqrt{g}N^{-D/z}\mG\equiv \Phi_1
\approx 0 \ ,
\nonumber \\
\partial_t p^i&=&\pb{p^i,H_T}=
-\mH^i \approx 0 \ ,  \nonumber \\
\end{eqnarray}
where $\Phi_1\approx 0$ and
$\mH^i\approx 0$ are secondary
constraints.  More intricate is the
analysis of the preservation of the
constraint $\mD$. We start with the
observation that the Poisson bracket
between $\mD$ and $\bT_S(\xi)$ is
proportional to $\mD$ and hence it
vanishes on the constraint surface.
Further we show that the Poisson
bracket between $\mD_0$ and $\mH_T$
vanishes. To begin with we determine
the Poisson brackets between $\mD_0$
and all phase space variables
\begin{eqnarray}
\pb{\mD_0(\bx),\pi^{ij}(\by)}&=&2\pi^{ij}(\bx)
\delta(\bx-\by) \ , \quad
\pb{\mD_0(\bx),g_{ij}(\by)}=-2
g_{ij}(\bx)\delta(\bx-\by) \ ,
\nonumber \\
\pb{\mD_0(\bx),N(\by)}&=&-zN(\bx)
\delta(\bx-\by) \ , \quad
\pb{\mD_0(\bx),p_N(\by)}= zp_N(\bx)
\delta(\bx-\by) \ . \nonumber \\
\end{eqnarray}
Then it is easy to see that
\begin{equation}
\pb{\mD_0(\bx),\bg_{ij}(\by)}=0
\end{equation}
that implies
\begin{eqnarray}
\pb{\mD_0(\bx),\bR(\by)}=0
\ , \quad
\pb{\mD_0(\bx),
\sqrt{g(\by)}}=
-D\sqrt{g(\bx)}\delta(\bx-\by) \ .
\nonumber \\
\end{eqnarray}
Using this result  together with
\begin{eqnarray}
\pb{\mD_0(\bx),\frac{N^{D/z}}{\sqrt{g}}
\pi^{ij}\mG_{ijkl}\pi^{kl}(\by)}=0
\nonumber \\
\end{eqnarray}
we find
\begin{eqnarray}
\pb{\mD_0(\bx),\mH_T(\by)}=0 \ .
\nonumber \\
\end{eqnarray}
Now we proceed to the analysis of the
time evolution of the second term in
$\mD$ that is proportional to  $
\mG(\bR)$. We use the fact that
\begin{eqnarray}
\pb{\mG(\bR)(\bx),\pi^{ij}(\by)}=
\frac{d\mG}{d\bR}\pb{\bR(\bx),\pi^{ij}(\by)}=
\frac{d\mG}{d\bR}(\bx) \frac{\delta
\bR(\bx)} {\delta \bg_{ij}(\by)}
\frac{1}{N^{2/z}(\by)} \ .
\nonumber \\
\end{eqnarray}
Then with the help of formulas
\begin{eqnarray}
& &\frac{\delta \bR(\bx)}{\delta
\bg_{ij}(\by)}=
-\bR^{ij}(\bx)\delta(\bx-\by)+
\bar{\nabla}^i \bar{\nabla}^j
\delta(\bx-\by)-\bg^{ij}(\bx)
\bar{\nabla}_k
\bar{\nabla}^k\delta(\bx-\by) \ ,
\nonumber \\
& &\bnabla^i \bnabla^j \mG_{ijkl}
\pi^{kl}- \bg^{ij}\bnabla_m\bnabla^m
\bar{\mG}_{ijkl}\pi^{kl} =\bnabla_k (\bnabla_l
\pi^{kl}) +\frac{1-\lambda}{\lambda
D-1}\bnabla_i
\bnabla^i \pi \  \nonumber \\
\end{eqnarray}
we determine following Poisson bracket
\begin{eqnarray}
& &\pb{\mG(\bR)(\bx),\pi^{ij}(\by)}=
\frac{d\mG}{d\bR}(\bx)
\frac{1}{N^{2/z}(\by)} \times
\nonumber \\
&\times& (-\bR^{ij}(\bx)\delta(\bx-\by)+
\bar{\nabla}^i \bar{\nabla}^j
\delta(\bx-\by)-\bg^{ij}(\bx)
\bar{\nabla}_k
\bar{\nabla}^k\delta(\bx-\by)) \ .
\nonumber \\
\end{eqnarray}
Collecting all these results together
 we determine the time evolution of the
constraint $\mD$
\begin{eqnarray}
\partial_t\mD=\pb{\mD,H_T}\approx
 2\frac{d\mG}{d\bR}\Phi_\mD^{II}\approx 0
\nonumber \\
\end{eqnarray}
where
\begin{equation}
\Phi_\mD^{II}= -
\bR^{ij}\bar{\mG}_{ijkl}N^{2/z}\pi^{kl}+
\bnabla_i\bnabla_j[N^{2/z}\pi^{ij}]
+\frac{1-\lambda}{D\lambda-1}
\bnabla_i\bnabla^i [N^{2/z}\bar{\pi}] \
.
\end{equation}
Note also that using definition of  $\mD$ and $\Phi_1$
we can introduce $\mD_0$ as an independent constraint
\begin{equation}
\mD_0=\mD-\Phi_1\approx 0 \ .
\end{equation}
In summary, we have following set of constraints
\begin{equation}
p^i\approx 0 \ , \quad  p_\mA\approx 0
\ , \quad \mH^i\approx 0 \ , \quad
\mD_0\approx 0 \ , \quad \Phi_1 \approx
0 \ , \quad \Phi_\mD^{II}\approx 0 \ .
\end{equation}
Now we proceed to the analysis of
 the stability of all
constraints taking into account an
existence of the secondary constraints.
This is simple task in case of  $p^i$
and $p_\mA$ that are trivially
preserved and form the first class
constraints of the theory. In the same
way we can show that $\bT_S(N)$ is
generator of the spatial diffeomorphism
and it is preserved as well.
 Let us now analyze
briefly the  constraint $\mD_0$. Since
$\pb{\mD_0(\bx),\bg_{ij}(\by)}=0$ and
$\pb{\mD_0(\bx),N^{z/2}\pi^{ij}(\by)}=0$
we see that Poisson brackets between
$\mD_0$ and $\Phi_1\approx
0,\Phi_\mD^{II}\approx 0$ vanish as
well. In other words $\mD_0$ is the
first class constraint and its smeared
form is the generator of the scaling
transformation.

Finally we have to determine the time
evolution of the constraint $\Phi_1$
and $\Phi_{\mD}^{II}$. To do this we
firstly show that
\begin{eqnarray}
\pb{\Phi_1(\bx),\Phi_\mD^{II}(\by)}\approx
 \mathbf{\triangle}(\bR,\bR_{ij},\bx,\by)+
\frac{(1-\lambda)(1-D)}{D\lambda-1}
\bnabla_i\bnabla^i\bnabla_j\bnabla^j
\delta(\bx-\by) \ ,  \nonumber \\
\end{eqnarray}
where $\mathbf{\triangle}$ is
complicated expression containing
covariant derivatives of delta function
which is also proportional to
$\bR_{ij}$ and $\bR$. On the other hand
the last expression is proportional to
the ordinary function and it vanishes
for $\lambda=1$. We see that the
Poisson bracket between $\Phi_1$ and
$\Phi^{II}_\mD$ is non-zero and depends
on the phase space variables. It is
important to stress that for
$\lambda\neq 1$ this Poisson bracket is
non-zero over all phase space. An
exceptional situation occurs for
$\lambda=1$ when we find that this
Poisson bracket vanishes for the
subspace of the phase space where
$\bR_{ij}=0$. We can analyze this
situation as in
\cite{Kluson:2010zn,Kluson:2011xx} with
the result that
 the theory on the subspace
$\bR_{ij}=0$ is effectively topological.

Excluding this exceptional case we see
that the Poisson bracket between $\Phi_1$
and $\Phi_\mD^{II}$ is non-zero so that they
are the second class constraints. As a result
we can easily determine the time evolution
of the constrains $\Phi_1 $ and $\Phi_\mD^{II}$.
In case of $\Phi_1$ we find
\begin{eqnarray}
\partial_t \Phi_1&=&\pb{\Phi_1,H_T}\approx
2\frac{d\mG}{d\bR}\Phi_{\mD}^{II}+\int d^D\bx
v_{II}^\mD \pb{\Phi_1,\Phi_\mD^{II}(\bx)}\approx
\nonumber \\
&\approx & \int d^D\bx v_{II}^\mD
\pb{\Phi_1,\Phi_\mD^{II}(\bx)}
=0 \nonumber \\
\end{eqnarray}
that implies $v_{\mD}^{II}=0$. On the
other hand the time evolution of the
constraint $\Phi_\mD^{II}$ is equal to
\begin{eqnarray}
\partial_t\Phi_\mD^{II}=
\pb{\Phi_\mD^{II},H_T}\approx \int
d^D\bx
\left(\pb{\Phi_\mD^{II},\mH_T(\bx)}\right.+
\nonumber \\
\left.+v_1\pb{\Phi_\mD^{II},\Phi_1(\bx)}+
v_\mD^{II}\pb{\Phi_\mD^{II},
\Phi_\mD^{II}(\bx)}\right)=0 \ .
\nonumber \\
\end{eqnarray}
Now due to the fact that $v_\mD^{II}=0$
we see that this equation determines
$v_1$ as a functional of canonical
variables, at least in principle. Say
differently, we see that it is not
necessary to impose additional
constraints on the systems and hence we
can stop here.

In summary,  the constraint structure
of given theory is the same as the constraint
structure of non-relativistic covariant
HL gravity studied in \cite{Kluson:2010zn}.
 This should not be surprising when we note that these
two theories differ in the form of the
St\"{u}ckelberg fields while the form of
the additional term is the same.
Explicitly, $\Phi_1$ and
$\Phi_\mD^{II}$ are the second
class constraints that,
according to the  standard
analysis have to vanish strongly
and hence they  allow us to express
two phase space variables as functions
of remaining physical phase space
variables, at least in principle.
Even if we cannot solve these constraints
explicitly in general case we can still
determine the number of physical degrees
of freedom. To do this note that there are
$D(D+1)$  gravity phase   space variables   $g_{ij}, \pi^{ij}$, $2D$
variables $N_i,p^i$,
$2$ variables $\mA,p_\mA$
and $2$  variables $N,p_N$.
In summary the total number of degrees of freedom is
 $N_{D.o.f}=D^2+3D+4$.
 On the other
hand we have $D$  first class
constraints $\mH^i\approx 0$, $D$ first
class constraints  $p^i\approx 0 $, $2$
first class constraints $\mD_0\approx 0
,p_\mA\approx 0$ and two second class
constraints $\Phi_1, \Phi_\mD^{II}$.
Then we have $N_{f.c.c}=2D+2$
  first class constraints and
 $N_{s.c.c.}=2$  second
class constraints. Then the
number of physical degrees of freedom
is \cite{Henneaux:1992ig}
\begin{equation}\label{pdf}
N_{D.o.f.}-2N_{f.c.c}-N_{s.c.c.}=
D^2-D-2 \
\end{equation}
that exactly corresponds to the number
of the phase space  physical degrees of freedom of $D+1$ dimensional
gravity.

As we argued previously in
\cite{Kluson:2010zn,Kluson:2011xx} it
is difficult to make further analysis
of these  second class constraints. For
example, the symplectic structure
defined by corresponding Dirac brackets
is very complicated. Secondly, it is
also subtle point to perform
quantization of the theory with the
second class constraints. In some
situations it is useful to perform so
named \emph{Abelian conversion}
\cite{Batalin:1991jm}. In this process
we extend given theory so that it
becomes  theory with the first class
constraints only. Even if this
procedure is possible in principle it
is again very difficult to apply it in
our case due to the complicated form of
the Poisson bracket between the second
class constraints. For these reasons
the proper understanding Weyl invariant
HL gravity is lacking.

 \noindent {\bf
Acknowledgements:}
 This work   was
supported by the Czech Ministry of
Education under Contract No. MSM
0021622409. \vskip 5mm



\begin{thebibliography}{20}


%
%
%
%
%
%
%
%
%

\bibitem{Horava:2009uw}
  P.~Horava,
\emph{``Quantum Gravity at a Lifshitz
Point,''}
  Phys.\ Rev.\  D {\bf 79} (2009) 084008
  [arXiv:0901.3775 [hep-th]].


\bibitem{Horava:2008ih}
  P.~Horava,
\emph{``Membranes at Quantum
Criticality,''}
  JHEP {\bf 0903} (2009) 020
  [arXiv:0812.4287 [hep-th]].

\bibitem{Horava:2008jf}
  P.~Horava,
\emph{``Quantum Criticality and
Yang-Mills Gauge Theory,''}
  arXiv:0811.2217 [hep-th].



\bibitem{Horava:2011gd}
  P.~Horava,
\emph{``General Covariance in Gravity
at a Lifshitz Point,''}
  arXiv:1101.1081 [hep-th].

\bibitem{Padilla:2010ge}
  A.~Padilla,
  \emph{``The good, the bad and
   the ugly .... of Horava gravity,''}
  arXiv:1009.4074 [hep-th].

\bibitem{Mukohyama:2010xz}
  S.~Mukohyama,
\emph{``Horava-Lifshitz Cosmology: A
Review,''}
  arXiv:1007.5199 [hep-th].

\bibitem{Weinfurtner:2010hz}
  S.~Weinfurtner, T.~P.~Sotiriou and M.~Visser,
\emph{``Projectable Horava-Lifshitz
gravity in a nutshell,''}
  J.\ Phys.\ Conf.\ Ser.\  {\bf 222}, 012054 (2010)
  [arXiv:1002.0308 [gr-qc]].

\bibitem{Sotiriou:2010wn}
  T.~P.~Sotiriou,
\emph{``Horava-Lifshitz gravity: a
status report,''}
  arXiv:1010.3218 [hep-th].


\bibitem{Visser:2011mf}
  M.~Visser,
\emph{``Status of Horava gravity: A
personal perspective,''}

  [arXiv:1103.5587 [hep-th]].







\bibitem{Sotiriou:2009bx}
  T.~P.~Sotiriou, M.~Visser and S.~Weinfurtner,
\emph{``Quantum gravity without Lorentz
invariance,''}
  JHEP {\bf 0910} (2009) 033
  [arXiv:0905.2798 [hep-th]].





\bibitem{Nojiri:2009th}
  S.~Nojiri and S.~D.~Odintsov,
\emph{``Covariant Horava-like
renormalizable gravity and its FRW
cosmology,''}
  Phys.\ Rev.\  D {\bf 81} (2010) 043001
  [arXiv:0905.4213 [hep-th]].

\bibitem{Nojiri:2010tv}
  S.~Nojiri and S.~D.~Odintsov,
\emph{``A proposal for covariant
renormalizable field theory of
gravity,''}
  Phys.\ Lett.\  B {\bf 691} (2010) 60
  [arXiv:1004.3613 [hep-th]].








 \bibitem{Blas:2009yd}
   D.~Blas, O.~Pujolas and S.~Sibiryakov,
 \emph{``On the Extra Mode and
 Inconsistency of Horava Gravity,''}
   JHEP {\bf 0910} (2009) 029
   [arXiv:0906.3046 [hep-th]].


\bibitem{Blas:2009qj}
  D.~Blas, O.~Pujolas and S.~Sibiryakov,
\emph{``A healthy extension of Horava
gravity,''}
  arXiv:0909.3525 [hep-th].



\bibitem{Blas:2009ck}
  D.~Blas, O.~Pujolas and S.~Sibiryakov,
\emph{``Comment on `Strong coupling in
extended Horava-Lifshitz gravity',''}
  arXiv:0912.0550 [hep-th].




\bibitem{Blas:2010hb}
  D.~Blas, O.~Pujolas and S.~Sibiryakov,
\emph{``Models of non-relativistic
 quantum gravity: the good, the bad and the
healthy,''}
  arXiv:1007.3503 [hep-th].





\bibitem{Li:2009bg}
  M.~Li and Y.~Pang,
\emph{``A Trouble with
Ho\v{r}ava-Lifshitz Gravity,''}
  JHEP {\bf 0908} (2009) 015
  [arXiv:0905.2751 [hep-th]].


\bibitem{Henneaux:2009zb}
  M.~Henneaux, A.~Kleinschmidt and G.~L.~Gomez,
\emph{``A dynamical inconsistency of
Horava gravity,''}
  arXiv:0912.0399 [hep-th].



\bibitem{Kluson:2010xx}
  J.~Kluson,
\emph{``Note About Hamiltonian
Formalism of Modified $F(R)$
Ho\v{r}ava-Lifshitz Gravities and Their
Healthy Extension,''}
  Phys.\ Rev.\  D {\bf 82} (2010) 044004
  [arXiv:1002.4859 [hep-th]].

\bibitem{Kluson:2010nf}
  J.~Kluson,
\emph{``Note About Hamiltonian
Formalism of Healthy Extended
Horava-Lifshitz Gravity,''}
  JHEP {\bf 1007} (2010) 038
  [arXiv:1004.3428 [hep-th]].


\bibitem{Bellorin:2010te}
  J.~Bellorin and A.~Restuccia,
\emph{``Closure of the algebra of
constraints for a non-projectable
Ho\v{r}ava model,''}
  arXiv:1010.5531 [hep-th].


\bibitem{Bellorin:2010je}
  J.~Bellorin and A.~Restuccia,
\emph{``On the consistency of the
Horava Theory,''}
  arXiv:1004.0055 [hep-th].


\bibitem{Kobakhidze:2009zr}
  A.~Kobakhidze,
\emph{``On the infrared limit of
Horava's gravity with the global
Hamiltonian constraint,''}
  Phys.\ Rev.\  D {\bf 82} (2010) 064011
  [arXiv:0906.5401 [hep-th]].

\bibitem{Pons:2010ke}
  J.~M.~Pons and P.~Talavera,
  \emph{``Remarks on the consistency
   of minimal deviations from General Relativity,''}
  Phys.\ Rev.\  D {\bf 82} (2010) 044011
  [arXiv:1003.3811 [gr-qc]].






\bibitem{Kluson:2010na}
  J.~Kluson,
\emph{``Horava-Lifshitz Gravity And
Ghost Condensation,''}
  Phys.\ Rev.\  D {\bf 82} (2010) 124011
  [arXiv:1008.5297 [hep-th]].





%



\bibitem{Horava:2010zj}
  P.~Horava and C.~M.~Melby-Thompson,
\emph{``General Covariance
 in Quantum Gravity at a Lifshitz Point,''}
  Phys.\ Rev.\  D {\bf 82} (2010) 064027
  [arXiv:1007.2410 [hep-th]].


\bibitem{daSilva:2010bm}
  A.~M.~da Silva,
\emph{``An Alternative Approach for
General Covariant Horava-Lifshitz
Gravity and Matter Coupling,''}
  arXiv:1009.4885 [hep-th].




\bibitem{Greenwald:2010fp}
  J.~Greenwald, V.~H.~Satheeshkumar and A.~Wang,
\emph{``Black holes, compact objects
and solar system tests in
non-relativistic general covariant
theory of gravity,''}
  arXiv:1010.3794 [hep-th].

\bibitem{Alexandre:2010cb}
  J.~Alexandre and P.~Pasipoularides,
\emph{``Spherically symmetric solutions
in Covariant Horava-Lifshitz
Gravity,''}
  arXiv:1010.3634 [hep-th].

\bibitem{Wang:2010wi}
  A.~Wang and Y.~Wu,
\emph{``Cosmology in nonrelativistic
general covariant theory of gravity,''}
  arXiv:1009.2089 [hep-th].



\bibitem{Huang:2010ay}
  Y.~Huang and A.~Wang,
\emph{``Nonrelativistic general
covariant theory of gravity with a
running constant $\lambda$,''}
  arXiv:1011.0739 [hep-th].


\bibitem{Kluson:2010zn}
  J.~Kluson,
\emph{``Hamiltonian Analysis of
Non-Relativistic Covariant RFDiff
Horava-Lifshitz Gravity,''}
  Phys.\ Rev.\  {\bf D83}, 044049 (2011).
  [arXiv:1011.1857 [hep-th]].

\bibitem{Kluson:2010za}
  J.~Kluson, S.~Nojiri, S.~D.~Odintsov and D.~Saez-Gomez,
\emph{``U(1) Invariant F(R)
Horava-Lifshitz Gravity,''}
  arXiv:1012.0473 [hep-th].

\bibitem{Kluson:2011xx}
  J.~Kluson,
\emph{``Lagrange Multiplier Modified
Horava-Lifshitz Gravity,''}
  [arXiv:1101.5880 [hep-th]].


\bibitem{Ruegg:2003ps}
  H.~Ruegg, M.~Ruiz-Altaba,
\emph{``The Stuckelberg field,''}
  Int.\ J.\ Mod.\ Phys.\  {\bf A19}, 3265-3348 (2004).
  [hep-th/0304245].

\bibitem{Henneaux:1992ig}
  M.~Henneaux and C.~Teitelboim,
\emph{``Quantization of gauge
systems,''}
{\it  Princeton, USA: Univ. Pr. (1992)
520 p}

\bibitem{Wang:2010uga}
  A.~Wang, Q.~Wu,
\emph{``Stability of spin-0 graviton
and strong coupling in Horava-Lifshitz
theory of gravity,''}
  Phys.\ Rev.\  {\bf D83 } (2011)  044025.
  [arXiv:1009.0268 [hep-th]].


\bibitem{Moon:2010wq}
  T.~Moon, P.~Oh, J.~Sohn,
\emph{``Anisotropic Weyl symmetry and
cosmology,''}
  JCAP {\bf 1011}, 005 (2010).
  [arXiv:1002.2549 [hep-th]].

\bibitem{Bogdanos:2009uj}
  C.~Bogdanos, E.~N.~Saridakis,
\emph{``Perturbative instabilities in
Horava gravity,''}
  Class.\ Quant.\ Grav.\  {\bf 27 } (2010)  075005.
  [arXiv:0907.1636 [hep-th]].





\bibitem{Batalin:1991jm}
  I.~A.~Batalin and I.~V.~Tyutin,
\emph{``Existence theorem for the
effective gauge algebra in the
generalized
  canonical formalism with Abelian conversion
of second class constraints,''}
  Int.\ J.\ Mod.\ Phys.\  A {\bf 6} (1991) 3255.

\bibitem{Rubakov:2008nh}
  V.~A.~Rubakov and P.~G.~Tinyakov,
\emph{``Infrared-modified gravities and
massive gravitons,''}
  Phys.\ Usp.\  {\bf 51} (2008) 759
  [arXiv:0802.4379 [hep-th]].




\end{thebibliography}
\end{document}